\documentclass[conference]{IEEEtran}
\IEEEoverridecommandlockouts
\usepackage{cite}
\usepackage{amsmath,amssymb,amsfonts}
\usepackage{graphicx}
\usepackage{textcomp}
\usepackage{xcolor}
\usepackage{booktabs} 

\usepackage{enumerate}
\usepackage{enumitem}
\usepackage[capitalize]{cleveref}
\usepackage[caption=false,font=footnotesize]{subfig}
\usepackage[skip=2pt]{caption}
\usepackage{mdframed}
\usepackage{framed} 
\usepackage{bm}   
\usepackage{algorithm}
\usepackage{algpseudocode}
\usepackage{comment}

\usepackage{tikz}

\usepackage[
  paper=letterpaper,
  top=0.73in,
  bottom=1.02in,
  left=0.64in,
  right=0.64in,
  columnsep=0.25in
]{geometry}

\begin{document}
\title{Sensing-Assisted Anti-Blockage Pinching-Antenna Systems For Indoor Immersive Communications\vspace{-0.6cm}}

\author{
\IEEEauthorblockN{
Yi Hong\textsuperscript{1}, 
Yalin Liu\textsuperscript{1\dag},
Yulei Wang\textsuperscript{2},
Yaru Fu\textsuperscript{1}
}
\IEEEauthorblockA{
\textsuperscript{1}School of Science and Technology, Hong Kong Metropolitan University, Hong Kong SAR, China\\
\textsuperscript{2}College of Electronics and Information Engineering, South-Central Minzu University, Wuhan, China\\
}

\IEEEauthorblockA{Emails: \{yhong, ylliu, yfu\}@hkmu.edu.hk, ylwang@mail.scuec.edu.cn}
\thanks{This work was supported in part by the Research Grants Council of the Hong Kong Special Administrative Region, China, under Project UGC/FDS16/E15/24, and in part by the Hong Kong Metropolitan University Research Grant under Project PFDS/2025/34.}
\vspace{-1.1cm}
}

\maketitle

\begin{abstract}
Indoor immersive communication, empowered by millimeter-wave (mmWave) technologies, is a key enabler for 6G-ready Extended Reality (XR) applications. However, mmWave signals are highly susceptible to dynamic obstacles (e.g., moving humans), leading to frequent line-of-sight (LoS) blockages. Moreover, although massive-antenna systems have been widely adopted indoors, their fixed-site deployment limits their effectiveness in wide and dynamic indoor environments such as factories and offices. To address these challenges, we present a sensing-assisted anti-blockage Pinching-Antenna Systems (PASS) solution for indoor immersive communications. By mechanically moving a pinching antenna (PA) along a dielectric waveguide, PASS can flexibly extend the service coverage and re-establish LoS links in wide indoor scenarios. Besides, our solution integrates sensing capabilities into PASS, where the PA executes controlled micro-movements to transmit and receive chirp radar signals, enabling real-time sensing of obstacles. Based on the sensing results, we design an anti-blockage PA movement strategy that detects blocked regions along the waveguide and proactively moves the PA to safe zones. Accordingly, the whole system performs a cycled “sensing–movement–communication” workflow with each cycle completed in a limited frame to ensure both timely sensing and communication. To ensure system efficiency, a closed-form pre-optimized solution of sensing parameters is derived under practical mechanical and sensing constraints. Extensive experiments show that our solution reduces blockage-induced outages to near-zero while sustaining the effective throughput required by indoor immersive applications.
\end{abstract}


\section{Introduction}

Empowered by millimeter-wave (mmWave) technologies, indoor immersive communication is emerging as a key evolutionary trend toward 6G. It enables human-centric interactive services with the digital world to support various Extended Reality (XR) applications, such as smart healthcare, remote working, and XR gaming. A crucial requirement is to provide extremely high data-rate and low-latency services; for example, a typical XR scenario requires video frame rates of $90-120$ frames/second \cite{TVCG.2023.3247057}. To support this extreme capacity, massive antenna systems, such as Huawei's LampSite X series, have been adopted for indoor mmWave communications \cite{huawei2024tech96}. However, these massive antenna systems rely on fixed-site deployment, which strictly limits their service region, making it hard to cover wider indoor scenarios, e.g., factories and offices. Moreover, mmWave signals are highly susceptible to dynamic physical obstacles, such as humans moving randomly within a room \cite{6840343}. 
Considering the fixed-site deployment and mmWave links, these massive antenna systems cannot effectively bypass the dynamic obstacles, leading to severe line-of-sight (LoS) link blockages and communication outages \cite{glocom.2017.8254900}. To address the above issues, the emerging Pinching-Antenna Systems (PASS) paradigm provides a promising solution \cite{Liu2026Survey}. On one hand, by mechanically moving a pinching antenna (PA) along a dielectric waveguide \cite{liu2026pinching}, PASS can flexibly extend the service coverage and re-establish LoS links in wide indoor scenarios. On the other hand, the PA's inherent mechanical mobility enables anti-blockage capability to escape the geometrical shadow cast by physical obstacles \cite{11169486}. 

\begin{figure}[t]
\centering        
\includegraphics[width=7.3cm]{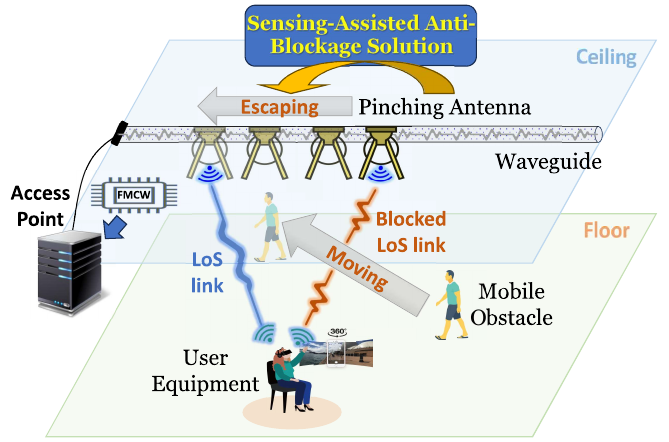}
\caption{An immersive communication scenario of our solution.}
\label{Fig-overview}
\end{figure}

Existing PASS studies present various PA anti-blockage solutions to enhance link quality \cite{10912473,11494673,11036558, 11315149, 11533561}. Foundational works explore flexible PASS solutions in various scenarios, e.g., NOMA-assisted activation \cite{10912473} and uplink rate maximization \cite{11494673}, but assume purely LoS links. Considering the LoS blockage, recent studies investigate different modeling methods of blockages to guide the PA's movement \cite{11036558, 11315149, 11533561}. By modeling obstacles as stochastic distributions, \cite{11036558} characterizes blockages for multi-user interference suppression and \cite{11315149} derives outage probabilities for multi-cell networks. Meanwhile, \cite{11533561} builds a three-dimensional (3D) geometry-aware model of eavesdropper-caused blockages and optimizes secure PASS operations \cite{11533561}. However, \cite{11036558, 11315149, 11533561} rely on long-term stochastic models, failing to provide real-time blockage characterization in the presence of dynamic obstacles. While the geometry-aware method in \cite{11533870} gives a deterministic blockage model, the considered obstacle (as an eavesdropper) is treated at a fixed point, which is insufficient for handling dynamic obstacles in indoor immersive communications. 

\begin{figure*}[t]
\centering        
\includegraphics[width=15cm]{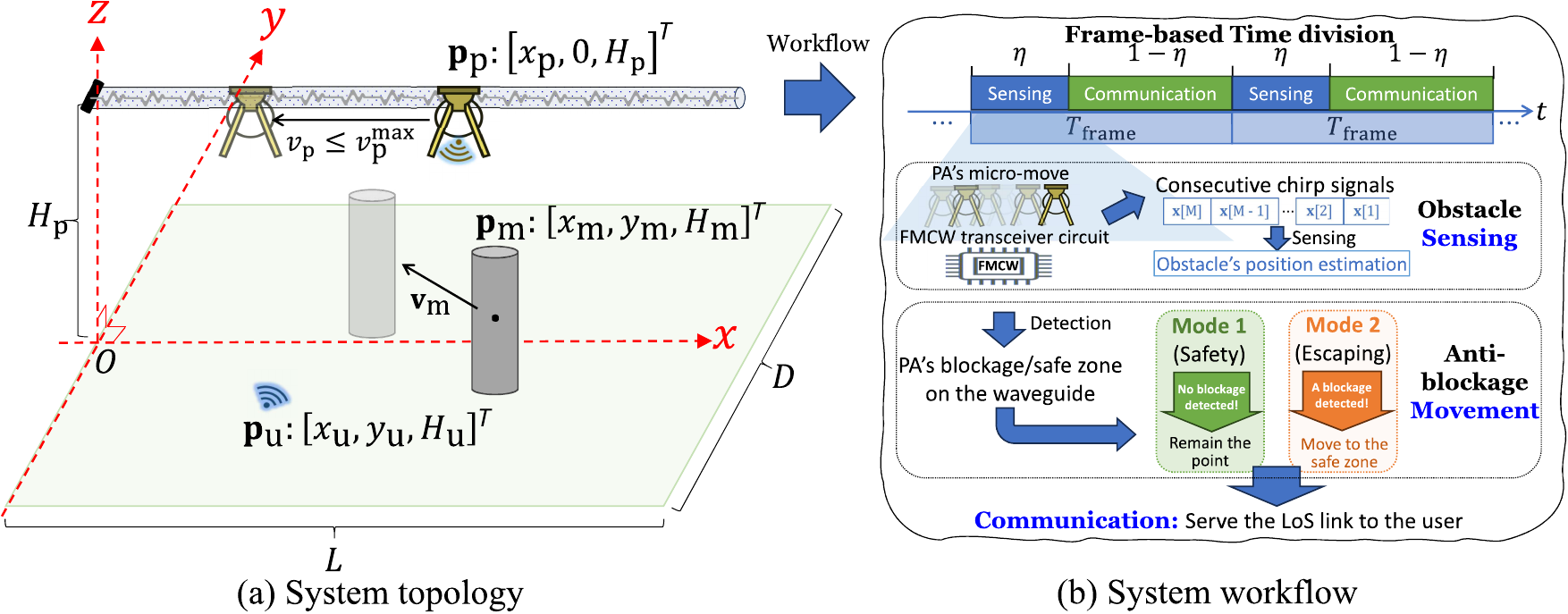}
\caption{A Sensing-Assisted Anti-Blockage PASS solution.}
\label{Fig-solution}
\end{figure*}

In applying PASS to indoor immersive communications, prior studies overlook a key practical limitation: the dynamic nature of obstacle positions. Obstacles can move randomly within the service area, resulting in time-varying and unknown locations that complicate the design of a PA movement strategy for anti-blockage. To address this issue, this paper proposes a sensing-assisted anti-blockage PASS solution (as shown in \cref{Fig-overview}). By integrating a Frequency-Modulated Continuous-Wave (FMCW) transceiver circuit into PASS, the system enables obstacle sensing based on ``chirp radar signals.'' Specifically, the PA performs controlled micro-movements to transmit and receive FMCW chirp signals, thereby sensing the obstacle’s kinematics in real time. Based on the sensing result, we design the anti-blockage PA movement strategy by detecting blocked boundaries of the PA over the waveguide and moving the PA to a safe zone. To support real-time data transmission, we develop a cycled ``sensing-movement-communication'' workflow, in which each cycle is completed within a bounded time to ensure timely sensing and dynamic PA movement. Meanwhile, to guarantee sensing efficiency, we derive a closed-form pre-optimized solution for the sensing parameters under the constraints imposed by the sensing requirements and the mechanical limits of the PA. Extensive experiments demonstrate that our solution suppresses blockage outages to near-zero levels while sustaining the effective throughput required for indoor immersive applications.

\section{System Configuration}
As shown in \cref{Fig-solution}, a Sensing-Assisted Anti-Blockage PASS solution is deployed in an indoor scenario, e.g., a room supporting immersive communication tasks. The PASS is employed by installing a dielectric waveguide on the ceiling with a PA moving along the waveguide to position its user. A mobile obstacle, e.g., a person, moves randomly within the room, thus causing link blockage to the PA-user communications. To address this issue, the solution performs periodical ``sensing-movement-communication'' cycles in continuous time frames. In each cycle, ``sensing'' is performed to estimate the location and motion of mobile obstacles, ``movement'' is moving the PA to a safe zone (``escaping'' from the obstacle), and ``communication'' is serving a stable LoS link for the user. 
To ensure operational efficiency, each cycle is completed within a time frame, denoted by $T_{\mathrm{frame}}$. $T_{\mathrm{frame}}$ is mainly allocated into the ``sensing'' and ``communication'' phases as the time in ``movement'' is comparatively small and ignorable. Let $\eta \in (0,1)$ be the frame ratio for sensing. Then $T_{\mathrm{frame}}$ is divided into: i) a sensing phase of duration $T_{\mathrm{s}}=\eta T_{\mathrm{frame}}$, and ii) a communication phase of duration $T_{\mathrm{c}}=(1-\eta)T_{\mathrm{frame}}$. Below are specific system configurations regarding system topology, sensing configurations, and signal propagation. The detailed workflow design is introduced in \cref{sec: workflow}. To ensure the effectiveness of the designed workflow, the feasible parameter settings are discussed in \cref{sec: para}.



\subsection{System Topology}
A PASS is deployed in an indoor room, where the service area is modeled as a three-dimensional (3D) rectangular space with length $L$, width $D$, and ceiling height $H_{\mathrm{p}}$. A dielectric waveguide of length $L$ is mounted centrally along the length of the ceiling at height $H_{\mathrm{p}}$. The user, located arbitrarily within the room, and the mobile obstacle are modeled as vertical cylinders with heights $H_{\mathrm{u}}$ and $H_{\mathrm{m}}$, respectively. The obstacle has a cross-sectional radius of $R_{\mathrm{m}}$.
A PA is driven by a mechanical motor to move along the waveguide. Due to the motor's physical limitation, the PA's movement is constrained by a maximum velocity $v_{\mathrm{p}}^{\max}$ and a physical acceleration $a_{\mathrm{p}}$. For simplicity, we ignore the motor's minor non-linear acceleration and deceleration transition states, approximating it as linear motion\footnote{A permanent magnet linear synchronous motor exhibits highly stable linear movement and ultra-rapid acceleration capabilities \cite{sato2015high}.}.
A 3D Cartesian coordinate system is established to locate all system objects, with its origin $O$ below the waveguide's feedpoint in the floor’s wider side (i.e., at the midpoint), the $xy$-plane on the floor, the $x$-axis opposite the waveguide feed points, the $y$-axis toward any direction perpendicular to the $x$-axis, and the $z$-axis vertically upward toward the ceiling’s midpoint. Let $\mathbf{p}_{\mathrm{p}}$, $\mathbf{p}_{\mathrm{u}}$, and $\mathbf{p}_{\mathrm{m}}$ be the coordinates of the PA, the user, and the geometrical centroid of the moving obstacle, respectively, which are given by
\begin{align*}
    &\mathbf{p}_{\mathrm{p}} : [x_{\mathrm{p}}, 0, H_{\mathrm{p}}]^{\mathsf{T}},
    \mathbf{p}_{\mathrm{u}} : [x_{\mathrm{u}}, y_{\mathrm{u}}, H_{\mathrm{u}}]^{\mathsf{T}},
    \mathbf{p}_{\mathrm{m}} : [x_{\mathrm{m}}, y_{\mathrm{m}}, H_{\mathrm{m}}]^{\mathsf{T}}.
\end{align*}%
Herein, $\{x_{\mathrm{u}},x_{\mathrm{m}}\}\in [0,L]$, $\{0,y_{\mathrm{u}},y_{\mathrm{m}}\}\in [-D/2, D/2]$ and $\{H_{\mathrm{u}},H_{\mathrm{m}}\} \in [0, H_{\mathrm{p}}]$ because $\mathbf{p}_{\mathrm{p}}$ is constrained by the waveguide while $\mathbf{p}_{\mathrm{u}}$ and $\mathbf{p}_{\mathrm{m}}$ locate/moves freely in the room.

\subsection{Sensing Configurations}
\label{subsubsec: sensing}
The PASS integrates an FMCW transceiver circuit in \cite{pyo2016single}, enabling the sensing capability via simultaneous chirp emission and echo collection through a single antenna. 
To position the mobile obstacle, the PA conducts a controlled micro-movement along the waveguide while simultaneously emitting and collecting a sequence of consecutive chirp signals. Particularly, during a sensing phase $T_{\mathrm{s}}$, the PA micro-moves to a direction via a scanning velocity $v_{\mathrm{scan}}$ ($v_{\mathrm{scan}}\leq v_{\mathrm{p}}^{\max}$). Let $f_{\mathrm{PRF}}$, $\Delta t_{\mathrm{s}}$, and $f_{\mathrm{s}}$, be the chirp repetition frequency of chirp signals, the duration of a single chirp, and the sampling rate $f_{\mathrm{s}}$ of emitting/collecting chirp signals. Thus, the echoing signals extract the features of the mobile obstacle in two dimensions: the chirp dimension comprising $M=f_{\mathrm{PRF}}T_{\mathrm{s}}$ consecutive chirps, and the sampling dimension comprising $N= f_{\mathrm{s}}\Delta t_{\mathrm{s}}$ signal samples collected within each chirp. These two dimensions represent the number of chirp signals emitted during $T_{\mathrm{s}}$ and the number of chirp-signal samples collected during $\Delta t_{\mathrm{s}}$, representing $N$ samples enclosed within each of the $M$ consecutive chirps.


Since the PA conducts a micro-movement on the waveguide at a scanning velocity $v_{scan}$, the $M$ chirps are transmitted and received at different positions. By combining these echoing signals, a Synthetic Linear Array (SLA) comprising $M$ antenna elements is synthesized \cite{6504845}. In the SLA, the interval between two adjacent virtual antenna elements is given by
\vspace{-0.2cm}

{\small
\begin{equation}
    \Delta d_{\mathrm{m}}=\frac{v_{\mathrm{scan}}T_{\mathrm{s}}}{M}=\frac{v_{\mathrm{scan}} T_{\mathrm{s}}}{f_{\mathrm{PRF}}T_{\mathrm{s}}}=\frac{v_{\mathrm{scan}}}{f_{\mathrm{PRF}}}.
\end{equation}}%

Accordingly, the effective synthetic aperture length of SLA is $L_{\mathrm{SA}} = M \Delta d_m = v_{\mathrm{scan}} T_{\mathrm{s}}$. To guarantee fundamental spatial discrimination, a minimum cross-range angular resolution (denoted $\Delta\theta_{\mathrm{req}}$) should be satisfied. Based on the Rayleigh criterion $\lambda_{\mathrm{s}} / (2 L_{\mathrm{SA}}) \le \sin(\Delta\theta_{\mathrm{req}})$~\cite{6504845}, a rigorous kinematic lower bound is explicitly imposed, which is given by
\vspace{-0.2cm}

{\small
\begin{equation}
    v_{\mathrm{scan}} = \frac{L_{\mathrm{SA}}}{T_{\mathrm{s}}} \ge \frac{\lambda_{\mathrm{s}}}{2 T_{\mathrm{s}} \sin(\Delta\theta_{\mathrm{req}})},
\end{equation}}%
where $\lambda_{\mathrm{s}}$ is the wavelength of sensing signals. 
\vspace{-0.2cm}

\subsection{Signal Propagation} Signals emitted from the PA are used for sensing and communication. 
These signals propagate along the waveguide and the PA. Let $G_{\mathrm{w}} = \rho^2 e^{-2\alpha_{\mathrm{wg}}x_{\mathrm{p}}}$ be the power gain along the waveguide\cite{inproceedings}, where $\rho$ denotes the amplitude coupling efficiency and $\alpha_{\mathrm{wg}}$ is the waveguide attenuation coefficient. During the sensing phase $T_{\mathrm{s}}$, the effective Signal-to-Noise Ratio (SNR) is deemed as the multiplied raw single-sample SNR 
by the coherent processing gain of $NM$ samples. And the PA collects a total of $NM$ discrete signal samples across the SLA. By adding amplitudes of these echo signal samples, it yields a processing gain equal to $NM$. Consequently, the effective SNR based on all sensing signals from the SLA is given by
\vspace{-0.2cm}

{\small
\begin{equation}
\begin{split}
    \mathsf{SNR}_{\mathrm{s}}    &
    =\frac{P_{\mathrm{p}} G_{\mathrm{w}}^2 \sigma_{\mathrm{rcs}} \lambda_{\mathrm{s}}^2 
    }{(4\pi)^3 \hat{d}_{\mathrm{s}}^4 \sigma_{\mathrm{g}}^2 } \times  N M,
\end{split}
\end{equation}}%
where 
\(P_{\mathrm{p}} G_{\mathrm{w}}^2 \lambda_{\mathrm{s}}^2 \sigma_{\mathrm{rcs}} / ((4\pi)^3 \hat{d}_{\mathrm{s}}^4)\) represents the backscattered signal power captured by the PA's effective synthetic aperture, $\sigma_{\mathrm{rcs}} = 2\pi R_{\mathrm{m}} H_{\mathrm{m}}^2 / \lambda_{\mathrm{s}}$ represents the effective radar cross-section of the mobile obstacle, $P_{\mathrm{p}}$ be the transmit power at the PA and $\hat{d}_{\mathrm{s}}$ is the estimated mobile obstacle distance. 

During the communication phase $T_{\mathrm{c}}$, the PA positions to the user and transmits data to the user via a LoS path. 
A unidirectional antenna is used for the user; thus, its antenna gain is 1. Let $\mathsf{SNR}_{\mathrm{c}}$ and $R_{\mathrm{c}}$ be the SNR and the communication rate of the PA-user link, which are given by
\vspace{-0.5cm}

{\small
\begin{align}
    &\mathsf{SNR}_{\mathrm{c}} = \frac{P_{\mathrm{p}}G_{\mathrm{w}} \lambda_{\mathrm{c}}^2 }{16\pi^2 d_{\mathrm{c}}^2\sigma_{\mathrm{g}}^2} , \\
    &R_{\mathrm{c}} = B_{\mathrm{c}} \log_2(1 + \mathsf{SNR}_{\mathrm{c}}),
    \label{Eq:Rate_c}
\end{align}}%
where $\lambda_{\mathrm{c}}$ is the wavelength of communication signals emitted from the PA, $B_{\mathrm{c}}$ is the used bandwidth, $\sigma_{\mathrm{g}}^2$ is the power of the Gaussian white noise $\mathcal{N}(0,\sigma_{\mathrm{g}}^2)$ at the receiver, and $d_{\mathrm{c}}$ is the distance between the PA and the user. 



\section{System Workflow}
\label{sec: workflow}

This section introduces the detailed design of the per-cycle workflow of ``sensing-movement-communication''. Since the communication is to serve the stable LoS with the user following the determined PA's movement, here we mainly introduce the sensing (for obstacle detection) and the PA movement (for anti-blockage once the obstacle is detected). 

\subsection{Obstacle Sensing}
\label{subsubsec: sensing_method}
The PA senses the obstacle via the 5 steps below: 

\subsubsection{\textsf{Step 1.1}} \textit{Sensing signal emitting, collecting, and transformation.} During a sensing phase $T_{\mathrm{s}}$, the PA micro-moves along the waveguide at a scanning speed $v_{\mathrm{scan}}$ and simultaneously emits $M$ continuous chirp signals. Based on the FMCW transceiver circuit in \cref{subsubsec: sensing}, the PA collects echo signals of $M$ chirp signals, denoted the echo matrix by $\mathbf{X} = [\mathbf{x}_1, \mathbf{x}_2, \dots, \mathbf{x}_M] \in \mathbb{C}^{N \times M}$, where the column vector $\mathbf{x}_m$ contains the $N$ complex signal samples collected within the $m$-th chirp. The PA's micro-moves synthesizes an SLA comprising $M$ discrete spatial elements.  
A Fast Fourier Transform (FFT) is applied to each column of matrix $\mathbf{X}$, converting the echo delay into a range spectrum comprising $N$ frequency bins, which are correlated with the distance of the mobile obstacle. A FFT is then executed across the $M$ chirps, transforming the shift induced by PA micro-movements into a feature matrix comprising $M$ angle bins, which correspond to the Direction-of-Arrival (DoA) of the mobile obstacle. This two-dimensional FFT yields a Delay - DOA feature matrix\footnote{Real-time obstacle localization can be computationally guaranteed by the lightweight FFT. Given $N$ samples and $M$ chirps ($N \gg M$), the overall time complexity simplifies directly to $\mathcal{O}(M N \log N)$, ensuring that sensing and parameter estimation are efficiently resolved within the dedicated slot $T_{\mathrm{s}}$.}. For quantitative analysis, all $NM$ bins in the feature matrix are indexed by $i$. Let $\mathbf{Y}$ denote the Delay-DOA feature matrix, and $A_i$ denote the raw amplitude of the $i$-th bin. The signal transformation and amplitude extraction are expressed as
\begin{equation}
\begin{aligned}
    \mathbf{Y} &= \mathcal{F}(\mathbf{X}) \in \mathbb{C}^{N \times M}, 
    \mathbf{y} = \mathrm{vec}(\mathbf{Y}), \\
    A_i &= |[\mathbf{y}]_i|, i \in \{1, 2, \dots, NM\},
\end{aligned}
\end{equation}
where $\mathcal{F}(\cdot)$ is the FFT operation, $\mathrm{vec}(\cdot)$ is the vectorization operator, which transforms into a single column vector, and $|\cdot|$ calculates the complex magnitude, mapping the complex values to absolute energy intensities in the real domain. Each element $y_i$ in vector $\mathbf{y}$ is a complex value representing the reflected signal at a specific range and angle coordinate, and its magnitude $A_i$ represents the echo strength from an obstacle at this coordinate.


\subsubsection{\textsf{Step 1.2}} \textit{Echo signals selection.} The raw amplitudes $A_i$ are contaminated by hardware imperfections and propagation clutter, and direct peak extraction remains vulnerable to false alarms. Sensing the moving obstacle, i.e., finding the affected echo signal, requires a dynamic amplitude-thresholding mechanism. 
A median-based threshold method was designed to filter out background noise by sorting all raw amplitudes $A_i$ in descending order, i.e., $\mathrm{median}(A_i)$. Specifically, an echo signal $i$ is identified as being affected by the moving obstacle only if its peak amplitude $A_k$ exceeds the median of the amplitude matrix. Let $\mathcal{K}_{\mathrm{cand}}$ denote the set of identified candidate echo signals, which is determined by
\begin{align}
\mathcal{K}_{\mathrm{cand}} = \left\{ k \mid k\in[1,2,..,NM], A_k\ge \mathrm{median}(A_i) \right\}.
\end{align}

\subsubsection{\textsf{Step 1.3}} \textit{Geometrical obstacle coordinate estimation.} For the $k$-th amplitudes $A_k$ extracted from the candidate set $\mathcal{K}_{\mathrm{cand}}$, the matrix indices are mapped to the beat frequency $\hat{f}_{\mathrm{p},k}$ and the spatial frequency index $\hat{m}_{\mathrm{p},k}$, expressed as
\begin{align}
\hat{f}_{\mathrm{p},k} &= \hat{n}_k \frac{f_{\mathrm{s}}}{N}, 
\hat{m}_{\mathrm{p},k} = \hat{m}_k - \frac{M}{2},
\end{align}
where $(\hat{n}_k, \hat{m}_k)$ denotes its coordinate in the Distance-Angle feature matrix $\mathbf{Y}$. 
Based on the coordinates of each identified echo signal within $\mathcal{K}_{\mathrm{cand}}$ in the Delay - DOA feature matrix $\mathbf{Y}$, the spatial coordinates of the obstacle can be estimated. Specifically, the distance $\hat{d}_{\mathrm{s},k}$ between the PA and the obstacle and the direction of arrival (DoA) $\hat{\theta}_{\mathrm{s},k}$ of the signal reflected at the PA are estimated by~\cite[Eqs. 2.17 and 2.32]{suleymanov2016design}
\begin{align}
    \hat{d}_{\mathrm{s},k} &= \frac{c \Delta t_{\mathrm{s}} \hat{f}_{\mathrm{p},k}}{2B_{\mathrm{s}}},
    \hat{\theta}_{\mathrm{s},k}= \arcsin \left( \frac{\hat{m}_{\mathrm{p},k} \lambda_{\mathrm{s}}}{2M \Delta d_{\mathrm{m}}} \right) ,
    \label{eq_hattheta}
\end{align}
where $c$ is the light speed, $B_{\mathrm{s}}$ is the bandwidth of the sensing chirp signal, and $\lambda_{\mathrm{s}}$ is the wavelength of the sensing chirp signal. Given a rough height $H_{\mathrm{m}}$ of the mobile obstacle (as a system-level preliminary setting), its coordinates $\hat{\mathbf{p}}_k = [\hat{x}_k, \hat{y}_k, H_{\mathrm{m}}]^{\mathsf{T}}$ can be reconstructed. According to the geometrical relationship between the obstacle and the PA, $\hat{x}_k$ and $\hat{y}_k$ are estimated by 
\begin{align}
    \hat{x}_{k} &= x_\mathrm{p} + \hat{d}_{\mathrm{s},k} \cos \hat{\theta}_{\mathrm{s},k}, \label{eq:x_est} \\
    \hat{y}_{k} &= \sqrt{\hat{d}_{\mathrm{s},k}^{2} - (\hat{x}_{k} - x_\mathrm{p})^{2} - (H_\mathrm{p} - H_\mathrm{m})^{2}}.
\label{Eq:TargetPosition_y}
\end{align}

\subsubsection{\textsf{Step 1.4}} \textit{User Localization Based on Projection Error Compensation:} 
In order to exclude the interference of the user from the estimated coordinate of the mobile obstacle, we first obtain the user's prior 3D spatial position via channel state information (CSI) feedback. Let $[x_{\mathrm{u}}, y_{\mathrm{u}}, H_{\mathrm{u}}]^{\mathsf{T}}$ denote the user's true coordinate derived from CSI. It is worth noting that the radar strictly resolves the 2D position based on the presumed target plane $Z = H_m$. Since the user equipment operates at a different height ($H_u \neq H_m$), the user's echo suffers from an elevation-induced range distortion when projected onto the radar's presumed 2D plane. To accurately exclude the user, our system derives the user's apparent position $\hat{y}_{\mathrm{u}}$ on the presumed $H_m$ plane based on 3D slant-range conservation:
\begin{equation}
    \hat{y}_{\mathrm{u}} = \sqrt{y_{\mathrm{u}}^2 + (H_{\mathrm{p}} - H_{\mathrm{u}})^2 - (H_{\mathrm{p}} - H_{\mathrm{m}})^2}.
\end{equation}

Consequently, the calibrated 3D coordinate of the user is obtained as $\hat{\mathbf{p}}_{\mathrm{u}} = [x_{\mathrm{u}}, \hat{y}_{\mathrm{u}}, H_{\mathrm{u}}]^{\mathsf{T}}$.

\subsubsection{\textsf{Step 1.5}} \textit{User-interference exclusion and coordinate calibration.} With the calibrated apparent position $\hat{\mathbf{p}}_{\mathrm{u}}$, we can now accurately separate the mobile obstacle from the user. Let $\hat{\mathbf{p}}_k = [\hat{x}_k, \hat{y}_k]^{\mathsf{T}}$ denote the estimated horizontal position of the moving obstacle from the $k$-th echo signal ($k\in \mathcal{K}_{\mathrm{cand}}$). The reconstructed candidate set $\mathcal{K}_\mathbf{cand}$ simultaneously retains the amplitude $A_k$ of each optional point for both the mobile obstacle and the user. Let $R_{\mathrm{u}}$ be the radius of a circular blanking region established around $\hat{\mathbf{p}}_{\mathrm{u}}$. The final estimated coordinate $\hat{\mathbf{p}}_{\mathrm{m}}$ is determined by selecting the candidate with the maximum amplitude $A_k$ falling outside this radius, which is formulated by
\vspace{-0.3cm}

{\footnotesize
\begin{equation}
[\hat{x}_{\mathrm{m}}, \hat{y}_{\mathrm{m}}]^{\mathsf{T}}= \hat{\mathbf{p}}_{k^*}, k^*=\mathop{\arg\max}_{k\in \mathcal{K}_{\mathrm{cand}}} \left\{ A_k \ \middle \vert \  \left\| \hat{\mathbf{p}}_k - \hat{\mathbf{p}}_{\mathrm{u}} \right\|_2 > R_{\mathrm{u}} \right\}.
\label{eq_mobile obstacle_localization}
\end{equation}
}%
\noindent Coupled with the prior height $H_{\mathrm{m}}$, the full 3D coordinate $\hat{\mathbf{p}}_{\mathrm{m}}:[\hat{x}_{\mathrm{m}}, \hat{y}_{\mathrm{m}}, H_{\mathrm{m}}]^{\mathsf{T}}$ is obtained. 

\subsection{Anti-blockage Movement}
\label{subsec: movement}


Based on the sensing result, i.e., the moving obstacle's coordinate, the PA performs the anti-blockage strategy in two steps: \textsf{Step 2.1:} formulating the PA's movable boundaries (i.e., defining the safe zone and PA's kinematic reachability limits), and \textsf{Step 2.2:} boundary-based movement.

\subsubsection{\textsf{Step 2.1}} \textit{Formulating the boundaries.} Based on the mobile obstacle's position $\hat{\mathbf{p}}_{\mathrm{m}}:[\hat{x}_{\mathrm{m}}, \hat{y}_{\mathrm{m}}, H_{\mathrm{m}}]^{\mathsf{T}}$ and its modeled cylinder shape with a cross-sectional radius $R_{\mathrm{m}}$, the mobile obstacle casts a geometric shadow onto the waveguide from the perspective of the user $\mathbf{p}_{\mathrm{u}}$. This instantiates a spatial mapping relationship that categorizes the available waveguide track into three distinct topological zones: the physical blockage zone (denoted by $\mathcal{Z}_{\mathrm{b}}$), the robust blockage zone (denoted by $\tilde{\mathcal{Z}}_{\mathrm{b}}$), and the safe zone (denoted by $\mathcal{Z}_{\mathrm{s}}$). 

Systematically, the physical shadow $\mathcal{Z}_{\mathrm{b}}$ is first dilated into $\tilde{\mathcal{Z}}_{\mathrm{b}}$ to accommodate uncertainties, which in turn deterministically bounds the complementary safe zone $\mathcal{Z}_{\mathrm{s}}$ for the PA's position and relocation. To determine whether an anti-blockage movement is necessary, a 3D vertical clearance condition is established. In practical indoor environments, if the user is sufficiently tall or the mobile obstacle is relatively short, the LoS signal can propagate directly over the obstacle without communication disconnection. Let $H_{\mathrm{LoS}}$ denote the vertical height of the LoS link line between the PA and the user, evaluated at the mobile obstacle's depth $\hat{y}_{\mathrm{m}}$. Since the heights of the PA and the user are fixed, based on similar triangles, the height of the LoS link is derived as
\vspace{-0.2cm}

{\small
\begin{equation}
    H_{\mathrm{LoS}} = H_{\mathrm{p}} - \frac{|\hat{y}_{\mathrm{m}}|}{|y_{\mathrm{u}}|} (H_{\mathrm{p}} - H_{\mathrm{u}}).
\label{eq: H_LoS}
\end{equation}}%

Based on this height, the blockage detection branches into two distinct conditions. If $H_{\mathrm{LoS}} > H_{\mathrm{m}}$ (where $H_{\mathrm{m}}$ is the known maximum height of indoor obstacles), the LoS link safely bypasses the obstacle vertically. In this scenario, the blockage effect is fully neutralized, meaning no blockage zone exists ($\mathcal{Z}_{\mathrm{b}} = \emptyset$). Consequently, the entire waveguide track is deemed a safe zone, i.e., $\mathcal{Z}_{\mathrm{s}} = [0, L]$, and no further movement is required. Conversely, if $H_{\mathrm{LoS}} \le H_{\mathrm{m}}$, the LoS link is occluded, triggering the 2D planar projection strategy to locate the horizontal shadow boundaries. Geometrically, a physical blockage zone $\mathcal{Z}_{\mathrm{b}}$ is identified on the waveguide, representing the interval where the mobile obstacle completely occludes the PA-user link. The physical blockage zone $\mathcal{Z}_{\mathrm{b}}$, which represents the shadow of the mobile obstacle projected onto the waveguide from the user's perspective, is formulated as
\vspace{-0.3cm}

\begin{figure}[t]
	\centering
        \includegraphics[width=6.5cm]{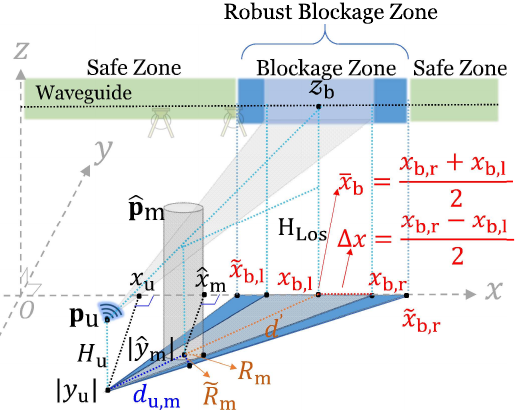}
    \caption{Geometrical relationship in blockage zone detection.}
    \label{Fig-collision range}
\end{figure}

{\small
\begin{equation}
    \mathcal{Z}_{\mathrm{b}} = 
    \begin{cases}
    \emptyset &,  \text{if } H_{\mathrm{LoS}} > H_{\mathrm{m}}, \\
    \left\{ [x_{\mathrm{b}}, 0, H_{\mathrm{p}}]^{\mathsf{T}} ,\forall x_{\mathrm{b}} \in [x_{\mathrm{b,l}}, x_{\mathrm{b,r}}] \right\} &,  \text{otherwise},
\end{cases}
\label{eq:blockage_zone_set}
\end{equation}}
with $x_{\mathrm{b,l}}$ and $x_{\mathrm{b,r}}$ corresponding to the left and right edge points of $\mathcal{Z}_{\mathrm{b}}$ from the perspective of the user, respectively. Let $d_{\mathrm{u,m}}$ be the distance between the user and the estimated mobile obstacle's position, which is given by $d_{\mathrm{u,m}} = \sqrt{(x_{\mathrm{u}} - \hat{x}_{\mathrm{m}})^2 + (y_{\mathrm{u}} - \hat{y}_{\mathrm{m}})^2}$. 
Let $\overline{x}_{\mathrm{b}}$ be the center of the blockage zone $\mathcal{Z}_{\mathrm{b}}$. According to the geometrical relationship in \cref{Fig-collision range}, since the user, the obstacle's center, and the shadow's center $\overline{x}_{\mathrm{b}}$ are strictly collinear, we formulate the equations based on similar triangles as follows
\vspace{-0.3cm}

{\small
\begin{align} 
    & \frac{\overline{x}_{\mathrm{b}}-\hat{x}_{\mathrm{m}}}{\left | \hat{y}_{\mathrm{m}} \right |} = \frac{\overline{x}_{\mathrm{b}}-{x}_{\mathrm{u}}}{\left | {y}_{\mathrm{u}} \right |}, \quad
    \frac{|\hat{y}_m|}{|y_{\mathrm{u}}|} = \frac{d'}{d' + d_{\mathrm{m,n}}}, \\
    & \frac{d_{\mathrm{m,n}}}{d_{\mathrm{m,n}}+d'} = \frac{R_{\mathrm{m}}}{\Delta x}, \\
    & \overline{x}_{\mathrm{b}}=\frac{x_{\mathrm{b,r}}+x_{\mathrm{b,l}}}{2} , \quad \Delta x=\frac{x_{\mathrm{b,r}}-x_{\mathrm{b,l}}}{2},
\end{align}
}%
where $d'$ is an auxiliary distance to assist our analysis. Solving the above equations of $x_{\mathrm{b,l}}$ and $x_{\mathrm{b,r}}$, we have
\vspace{-0.3cm}

{\small
\begin{align} 
    x_{\mathrm{b,l}} &= \frac{\hat{x}_{\mathrm{m}} |y_{\mathrm{u}}| - x_{\mathrm{u}} |\hat{y}_{\mathrm{m}}| - R_{\mathrm{m}} |y_{\mathrm{u}}|}{|y_{\mathrm{u}}| - |\hat{y}_{\mathrm{m}}|}, \notag \\
    x_{\mathrm{b,r}} &= \frac{\hat{x}_{\mathrm{m}} |y_{\mathrm{u}}| - x_{\mathrm{u}} |\hat{y}_{\mathrm{m}}| + R_{\mathrm{m}} |y_{\mathrm{u}}|}{|y_{\mathrm{u}}| - |\hat{y}_{\mathrm{m}}|},
    \notag \\|\mathcal{Z}_\mathrm{b}| &= x_{\mathrm{b,r}} - x_{\mathrm{b,l}}.
    \label{eq:blockage_zone}
\end{align}}%

To guarantee uninterrupted communication against sensing uncertainties, this ideal physical boundary is structurally expanded into a robust blockage zone $\tilde{\mathcal{Z}}_{\mathrm{b}}$. This region is constructed by dilating the exact physical radius $R_{\mathrm{m}}$ into a robust effective radius $\tilde{R}_{\mathrm{m}}$ to systematically compensate for sensing uncertainties, thereby guaranteeing uninterrupted communication. The accumulated spatial tracking error $\sigma_{\mathrm{m}}$, the robust radius $\tilde{R}_{\mathrm{m}}$, and the robust blockage zone $\tilde{\mathcal{Z}}_{\mathrm{b}}$ are formulated as
\vspace{-0.3cm}

{\footnotesize
\begin{align}
    \sigma_{\mathrm{m}} &= \sqrt{\sigma_{\mathrm{res}}^2 + \sigma_{\mathrm{glint}}^2 + \sigma_{\mathrm{SNR}}^2}, \label{eq_error_propagation} \\
    \tilde{R}_{\mathrm{m}} &= R_{\mathrm{m}} + \hat{v}_{\mathrm{m}} \left( \tau + \frac{1}{2} t_{\mathrm{outage}} \right) + \kappa\sigma_{\mathrm{m}}, \label{eq:robust_radius} \\
    \tilde{\mathcal{Z}}_{\mathrm{b}} &: \left\{ [x_{\mathrm{b}}, 0, H_{\mathrm{p}}]^{\mathsf{T}}  ,\forall x_{\mathrm{b}} \in [\tilde{x}_{\mathrm{b,l}}, \tilde{x}_{\mathrm{b,r}}] \right\}, \label{eq:robust_blockage_zone_set} \\
    |\tilde{\mathcal{Z}}_{\mathrm{b}}| &= \tilde{x}_{\mathrm{b,r}} - \tilde{x}_{\mathrm{b,l}} = \frac{2 \tilde{R}_{\mathrm{m}} |{y}_{\mathrm{u}}|}{|{y}_{\mathrm{u}}| - |\hat{y}_{\mathrm{m}}|},
    \label{eq:robust_blockage_zone_span}
\end{align}}%
where $\sigma_{\mathrm{res}} \approx c / (2B_{\mathrm{s}})$ is the bounded range measurement uncertainty, $\sigma_{\mathrm{glint}} \approx 2R_{\mathrm{m}} / \sqrt{12}$ is the theoretical Root Mean Square error caused by the mobile obstacle glint effect \footnote{The mobile obstacle is modeled as a cylinder, the radar reflection point drifts across its surface as the relative viewing angle changes during motion, which induces this spatial variance proportional to its physical dimensions.}, $\sigma_{\mathrm{SNR}} = \sqrt{ \left( \frac{c}{2B_{\mathrm{s}} \sqrt{\mathsf{SNR}_{\mathrm{s}}}} \right)^2 + \left( \frac{\hat{d}_{\mathrm{s}} \lambda_{\mathrm{s}}}{2 L_{\mathrm{SA}} \sqrt{\mathsf{SNR}_{\mathrm{s}}}} \right)^2 }$ denotes the thermal noise error, $\hat{v}_{\mathrm{m}} = \frac{\hat{\mathbf{p}}_{\mathrm{m}} - \hat{\mathbf{p}}_{\mathrm{m}}[t-1]}{T_{\mathrm{frame}}}$ is the prior velocity of the obstacle, where $t-1$ denotes the previous frame. and $\tau=v_{\mathrm{p}}^{\max}/a_{\mathrm{p}}$ denotes the mechanical actuation latency of the PA, $t_{\mathrm{outage}}$ represents the duration of sensing outages\footnote{PA misidentifies a mobile obstacle as an obstacle causing perceptual loss, and the tracker relies entirely on the previous frame's mobile obstacle movement speed to calculate the position.} times and $\kappa$ is a confidence multiplier.

Consequently, to maintain an uninterrupted communication link, the PA's movable area on the waveguide is strictly reduced to a safe zone $\mathcal{Z}_{\mathrm{s}}$. This secure region is derived by directly excluding the robust blockage interval from the entire waveguide length $[0, L]$. The exact spatial set of the safe zone and its total available spatial span, denoted by $|\mathcal{Z}_{\mathrm{s}}|$, are formulated as
\vspace{-0.3cm}

{\footnotesize
\begin{align}
\mathcal{Z}_{\mathrm{s}} &: \left\{ [x_{\mathrm{s}},0,H_{\mathrm{p}}]^{\mathsf{T}}, \forall x_{\mathrm{s}} \in [0, \tilde{x}_{\mathrm{b,l}}) \cup (\tilde{x}_{\mathrm{b,r}}, L] \right\}, \label{eq:safe_zone_set} \\
|\mathcal{Z}_{\mathrm{s}}| &= L - |\tilde{\mathcal{Z}}_{\mathrm{b}}| = L - \frac{2 \tilde{R}_{\mathrm{m}} |y_{\mathrm{u}}|}{|y_{\mathrm{u}}| - |\hat{y}_{\mathrm{m}}|}.
\label{eq:safe_zone_span}
\end{align}
}%

\subsubsection{\textsf{Step 2.1}} \textit{Boundary-Based Anti-blockage Movement.}
\label{subsec: evasion_strategy}

To execute the ``sensing-movement-communication'' cycle and maintain a LoS link, the PA must find an optimal position.  Assuming an ideal obstacle-free environment, the unconstrained optimal position maximizes the data ratio $R_{\mathrm{c}}(x_{\mathrm{p}})$ defined in \cref{Eq:Rate_c}. By incorporating the geometric spreading loss and denoting the $x_{\mathrm{p}}$-independent system constant as $C = P_{\mathrm{p}}\rho^{2}\lambda_{\mathrm{c}}^{2}/(16\pi^{2}\sigma_{\mathrm{g}}^{2})$, and the orthogonal distance between the user and the waveguide as $D_{0}^{2} = y_{\mathrm{u}}^{2} + (H_{\mathrm{p}} - H_{\mathrm{u}})^{2}$, the received SNR function is expressed as $\mathsf{SNR}_{\mathrm{c}}(x_{\mathrm{p}}) = Ce^{-2\alpha_{\mathrm{wg}}x_{\mathrm{p}}}/((x_{\mathrm{p}} - x_{\mathrm{u}})^{2} + D_{0}^{2})$. 

To find the unconstrained optimal position, we set the first-order derivative $\partial \mathsf{SNR}_{\mathrm{c}}/\partial x_{\mathrm{p}} = 0$, which reduces to finding the roots of the quadratic polynomial expressed as
\begin{equation}
\begin{split}
P(x_{\mathrm{p}}) &= \alpha_{\mathrm{wg}}x_{\mathrm{p}}^{2} + (1 - 2\alpha_{\mathrm{wg}}x_{\mathrm{u}})x_{\mathrm{p}} \\
&\quad + [\alpha_{\mathrm{wg}}(x_{\mathrm{u}}^{2} + D_{0}^{2}) - x_{\mathrm{u}}] = 0.
\end{split}
\end{equation}%

Since the discriminant $\Delta = 1 - 4\alpha_{\mathrm{wg}}^{2}D_{0}^{2}$ remains strictly positive ($4\alpha_{\mathrm{wg}}^2 D_0^2 \ll 1$ for typical indoor waveguides), solving $P(x_{\mathrm{p}}) = 0$ and verifying the negative second-order derivative yields the ideal unconstrained optimal position formulated as

\begin{equation}
x_{\mathrm{p}}^{\mathrm{opt}} = x_{\mathrm{u}} + \frac{-1 + \sqrt{1 - 4\alpha_{\mathrm{wg}}^{2}D_{0}^{2}}}{2\alpha_{\mathrm{wg}}}.
\label{Eq:OutputYc1}
\end{equation}

While \cref{Eq:OutputYc1} provides the unconstrained optimum position for maximizing the LoS link data ratio, this unconstrained position may be inside the blockage zone or exceed the PA's mobility capabilities. In practical deployment, the PA's position must be coupled with hardware kinematic limits to mitigate the impact of mechanical delays. Let $x_{\mathrm{p}, t-1}$ denote the PA's position in the previous frame $t-1$, and $x_{\mathrm{p}} \in \mathcal{Z}_{\mathrm{s}}$ denote the intended anchor of the PA for the current frame. To maintain uninterrupted communication, the maximum allowable mechanical displacement is bounded by the duration of the subsequent communication phase $T_{\mathrm{c}}$. Given the hardware's maximum operating velocity $v_{\mathrm{p}}^{\max}$, the PA's kinematic displacement is strictly bounded by $v_{\mathrm{p}}^{\max} (1 - \eta) T_{\mathrm{frame}}$. By coupling this kinematic reachability limit with the topological safe zone, we define a unified feasible set $\mathcal{X}_{\mathrm{safe}}$ as
\vspace{-0.1cm}

{\small
\begin{equation}
\mathcal{X}_{\mathrm{safe}} = \mathcal{Z}_{\mathrm{s}} \cap \left\{ x \mid |x - x_{\mathrm{p}, t-1}| \le v_{\mathrm{p}}^{\max} (1 - \eta) T_{\mathrm{frame}} \right\}.
\end{equation}
}%

To execute the ``sensing-movement-communication'' cycle and maintain an uninterrupted LoS link, the PA's anti-blockage strategy relies on the relationship between the ideal unconstrained optimal position $x_{\mathrm{p}}^{\mathrm{opt}}$ and the robust blockage zone $\tilde{\mathcal{Z}}_{\mathrm{b}}$. However, this ideal position $x_{\mathrm{p}}^{\mathrm{opt}}$ may violate $\tilde{\mathcal{Z}}_{\mathrm{b}}$ or the kinematic limits. The system categorizes the anti-blockage strategy into two operational modes to determine the actual constrained optimal position, denoted by $x_{\mathrm{p}}^*$:
\begin{itemize}[leftmargin=*,align=left]
    \item \textbf{Mode 1 (Safety):} If the unconstrained optimum position $x_{\mathrm{p}}^{\mathrm{opt}} \in \mathcal{X}_{\mathrm{safe}}$, the system assigns $x_{\mathrm{p}}^* = x_{\mathrm{p}}^{\mathrm{opt}}$. The PA operates normally without link blockage.
    \item \textbf{Mode 2 (Escaping):} If $x_{\mathrm{p}}^{\mathrm{opt}}$ within the robust blockage zone $\tilde{\mathcal{Z}}_{\mathrm{b}}$, the PA evades the obstacle (``escaping''), and this escaping process is driven by constraint evaluations. If a local feasible subspace remains on the current side of the user, the constrained optimum shifts to the boundaries. Consequently, the system evaluates the throughput at the valid extreme edges of $\mathcal{X}_{\mathrm{safe}}$ on this local side to determine $x_{\mathrm{p}}^*$. If $\tilde{\mathcal{Z}}_{\mathrm{b}}$ completely occupies the feasible interval on the current side, local boundary evaluation becomes invalid. The PA must cross the robust blockage zone $\tilde{\mathcal{Z}}_{\mathrm{b}}$ to reach the new position $x_{\mathrm{p}}^* \in \mathcal{X}_{\mathrm{safe}}$ on the opposite side.
\end{itemize}

By evaluating these mode-triggering conditions, the final constrained optimal position $x_{\mathrm{p}}^* \in \mathcal{X}_{\mathrm{mode}}$ is deterministically resolved in closed-form prior to system resource allocation.


\section{Pre-optimized Settings of Sensing Parameters}
\label{sec: para}

This section exploits the pre-optimized settings of several sensing parameters, including the sensing time ratio $\eta$ per frame and the PA scanning velocity $v_{\mathrm{scan}}$ during sensing. To guarantee the quality of communications, an upper bound on $\eta$ and its feasible value range are introduced. To improve sensing efficiency, a closed-form pre-optimized solution of $\eta$ and $v_{\mathrm{scan}}$ is derived under the constraints imposed by the sensing requirement and the mechanical limit of the PA. The derived solution can be adopted in our system before it runs.   

\subsection{Upper-bound of $\eta$}
Using the determined anti-blockage movement location in \cref{subsec: movement}, the PA serves communications to the user during the phase $T_{\mathrm{c}}=(1-\eta)T_{\mathrm{frame}}$. The effective data ratio over the whole frame $T_{\mathrm{frame}}$ should satisfy a QoS threshold $R_{\mathrm{QoS}}$ catering to a specific immersive application. Thus, to ensure a reliable connection, the effective throughput must satisfy the link quality constraint given by
\begin{equation}
    R_{\mathrm{eff}}=(1-\eta) \cdot R_{\mathrm{c}}(x_{\mathrm{p}}^*) \ge R_{\mathrm{QoS}}.\label{eq:QoS}
\end{equation}

The entire waveguide satisfies \cref{eq:QoS} under unobstructed conditions (Mode 1). 
However, intruding into the robust blockage zone degrades the link, thereby acting as the feasibility threshold that triggers the anti-blockage movement (Mode 2). To hold this constraint over all frames, the worst feasible PA-user link in our system should reach the above quality threshold. That is, when $\eta$ reaches a hard upper boundary $\eta_{\max}$ and $d_{\mathrm{c}}(x_{\mathrm{p}}^*)=\left | \mathbf{p}_{\mathrm{p}}(x_{\mathrm{p}}^*) - \mathbf{p}_{\mathrm{u}}\right | = d_{\mathrm{c}}^{\max}$ reaches the maximum tolerable link distance. Thus, we can derive $\eta_{\max}$ by
\begin{equation}
    \eta_{\max}\leq 1- \frac{R_{\mathrm{QoS}}}{R_{\mathrm{c}}(d_{\mathrm{c}}^{\max})}.\label{eq:eta}
\end{equation}

In the whole workflow of our solution, the value of $\eta_{\max}$ is set to ensure \cref{eq:eta}. Thus, \cref{eq:QoS} always holds, and the communication quality can be ensured. 

\subsection{Sensing Parameter Pre-optimization}
Since the link quality constraint always holds by using the upper boundary $\eta_{\max}$ in \cref{eq:QoS}, the remaining effective setting task is no longer to maximize the throughput, but to minimize the sensing duration $T_{\mathrm{s}}=\eta T_{\mathrm{frame}}$ under the feasible system constraints. Given a fixed $T_{\mathrm{frame}}$, the sensing duration minimization can be transformed to the sensing time ratio $\eta$ minimization. Therefore, two sensing parameters, i.e., the micro-motion velocity $v_{\mathrm{scan}}$ and the sensing duty cycle $\eta$ can be pre-optimized when configuring the system. The parameter pre-optimization is formulated as problem $\mathcal{P}1$.
\begin{align}
    \mathcal{P}1: \quad \min_{v_{\mathrm{scan}}, \eta} \quad & \eta, \notag \\
    \text{s.t.} \quad 
    & v_{\mathrm{scan}} \cdot \eta \ge \frac{\lambda_{\mathrm{s}}}{2T_{\mathrm{frame}}\sin(\Delta\theta_{\mathrm{req}})}, \tag{C1} \\
    & v_{\mathrm{scan}} \le v_{\mathrm{p}}^{\max}, \tag{C2} \\
    & \eta \le \eta_{\max}. \tag{C3}
\end{align}

In \(\mathcal{P}1\), (C1) defines the lower bound for the sensing resolution to distinguish the mobile obstacle from background clutter; (C2) restricts the sensing micro-motion velocity within the mechanical motor limit; and (C3) bounds the feasible sensing duty cycle overhead.

According to (C1), to satisfy the preset radar angular resolution, the sensing duty cycle must satisfy $\eta \ge \lambda_{\mathrm{s}}/(2T_{\mathrm{frame}}\sin(\Delta\theta_{\mathrm{req}}) v_{\mathrm{scan}})$. This lower bound monotonically decreases with respect to the micro-motion velocity $v_{\mathrm{scan}}$. Therefore, to minimize the temporal sensing overhead $\eta$, the system must operate the micro-motion velocity at its ceiling. Substituting the boundary of (C2) directly yields the optimal sensing velocity expressed as 



\begin{equation}
    v_{\mathrm{scan}}^{*} = v_{\mathrm{p}}^{\max},\eta^{*} = \frac{\lambda_{\mathrm{s}}}{2T_{\mathrm{frame}}\sin(\Delta\theta_{\mathrm{req}}) v_{\mathrm{p}}^{\max}}.
\end{equation}

Considering the hardware timing constraints (C3), the final optimal sensing duty cycle is bounded as

\begin{equation}
\eta^* = \min\left( \frac{\lambda_{\mathrm{s}}}{2T_{\mathrm{frame}}\sin(\Delta\theta_{\mathrm{req}}) v_{\mathrm{p}}^{\max}}, \eta_{\max}\right).
\end{equation}

This operation ensures that the system satisfies both the kinematic reachability limits of the mechanical motor and the required radar angular resolution, keeping the sensing duty cycle overhead to an absolute minimum. Consequently, the optimization problem $\mathcal{P}1$ is solved in closed-form. By transforming the coupled problem into geometric constraint evaluations and deterministic operational modes, the solution circumvents iterative numerical algorithms. This closed-form execution minimizes the online processing burden, ensuring the system executes the anti-blockage strategy with low latency for dynamic indoor immersive communication networks.




\section{Numerical Results}
This section conducts numerical simulation experiments to validate the performance of the developed sensing-assisted anti-blockage PASS solution. In our simulation, the rectangular room is considered with physical dimensions of $L \times D \times H_{\mathrm{p}} = 10.0 \times 10.0 \times 3.0 \text{ m}^3$. The dielectric waveguide spans the entire length $L$ along the ceiling's center axis ($y=0$). The user equipment is anchored at a fixed depth of $y_{\mathrm{u}} = 5$~m and a seated height of $H_{\mathrm{u}} = 0.9$~m, simulating a user engaging in an immersive gaming scenario. The user's $x$-coordinate varies within the room, acting as a system-level parameter to be analyzed in our experiments. The mobile obstacle is assumed as a standing adult pedestrian with a cross-sectional radius of $R_{\mathrm{m}} = 0.25$ m and a height of $H_{\mathrm{m}} = 1.8$ m, and a $y$-axis depth of $y_{\mathrm{m}} = 3.1$~m. It has the potential to meet the occlusion condition ($H_{\mathrm{LoS}} \le H_{\mathrm{m}}$) to trigger blockages for the PA-user communication. The detailed obstacle's moving trajectory is simulated and demonstrated in \cref{subsec: trajectory}. 
The primary metric for system performance is the effective throughput $R_{\mathrm{eff}}$. The system configurations are strictly aligned with 5G/6G mmWave standards, immersive communication service requirements, and practical hardware constraints. Specifically, the system parameters are set below.

\begin{figure}[t]
	\centering
        \includegraphics[width=8.8cm]{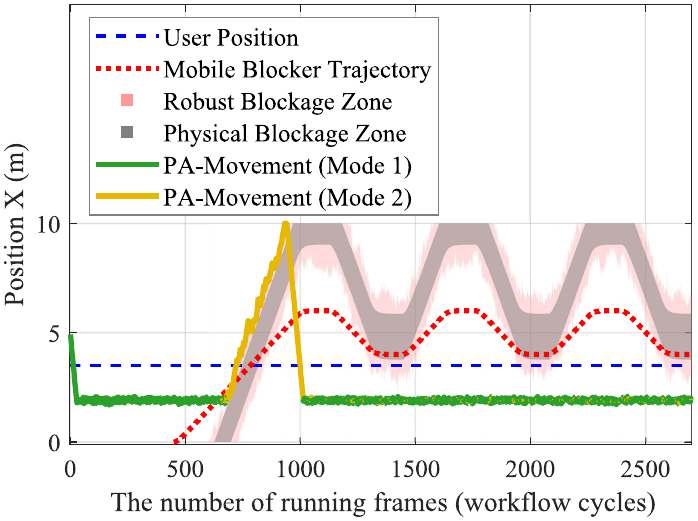}
    \caption{The PA's trajectory evolution over 30 seconds (a total of 2702 frames) following the anti-blockage movement modes. 
    }
    \label{Fig-experiment1-results}
\end{figure}

\begin{table*}[t]
\renewcommand{\arraystretch}{1.3} 
\centering
\caption{Structures of four compared baselines (including ours, CA1, CA2, and CA3)}
\label{table_structures}
\footnotesize 
\begin{tabular}{p{2cm}cccc}
\toprule
\textbf{Algorithms} & \textbf{Our Solution} & \textbf{\textit{CA1} - Static Physical Blockage} & \textbf{\textit{CA2} - Fixed Sensing} & \textbf{\textit{CA3} - Fixed PA} \\
\midrule
\textit{Effective Radius} & $\tilde{R}_m = R_m + \|\hat{v}_k\|\tau + \kappa\sigma_m$ [19] & $R_m$ & $\tilde{R}_m = R_m + \|\hat{v}_k\|\tau + \kappa\sigma_m$ [19] & N/A \\

\textit{Sensing Ratio} & $\eta^*$ (Optimal) [25] & $\eta^*$ (Optimal) [25] & $\eta = \eta_{\max}$ & $\eta = 0$ \\

\textit{PA Position} & Eq. (27) mode optimization $x_{\mathrm{p}}^*$ & Eq. (27) mode optimization $x_{\mathrm{p}}^*$ 
& Eq. (27) mode optimization $x_{\mathrm{p}}^*$ & Fixed $x_p = 5.0 \text{ m}$ \\
\bottomrule
\end{tabular}
\end{table*}

\begin{itemize}[leftmargin=*, nosep]
\item \textit{Communication Settings:} The system operates at a mmWave carrier frequency of $f_c = 28$ GHz\cite{11197530}, corresponding to a wavelength of $\lambda_{\mathrm{s}} = \lambda_{\mathrm{c}} \approx 10.71 \text{ mm}$, and the bandwidth of $B_{\mathrm{c}} = 400$ MHz. Physical transmission parameters include an amplitude coupling coefficient $\rho = 0.89$, waveguide attenuation $\alpha_{\mathrm{wg}} = 0.05 \text{ m}^{-1}$, receiver noise floor $\sigma_{\mathrm{g}}^2 = -80$ dBm, and transmit power $P_{\mathrm{p}} = 23$ dBm (following 3GPP TS 38.101-2 Power Class 3). To support uncompressed XR streaming at a 90 Hz refresh rate, the minimum QoS requirement is $R_{\mathrm{QoS}} = 500$ Mbps with a synchronized frame length of $T_{\mathrm{frame}} = 11.1$ ms. 

\item \textit{Sensing Settings:} The sensing bandwidth is $B_{\mathrm{s}} = 2$ GHz with a sensing antenna gain of 12 dBi. The chirp configuration includes a sampling rate $f_{\mathrm{s}} = 40$ MHz, $N = 1024$ samples per chirp, and a PRF of $20$ kHz. The mobile obstacle's radar cross-section is modeled as $\sigma_{\mathrm{rcs}} = 0.4 \text{ m}^2$ \cite{skolnik2008radar}. Aligning with 3GPP flexible symbol limits, the maximum sensing duty cycle is bounded at $\eta_{\max} = 25\%$, Under the defined room dimensions, this $\eta_{\max}$ satisfies the worst-case link condition in \eqref{eq:eta} at the maximum tolerable link distance $d_{\mathrm{c}}^{\max}$, continuously guaranteeing the required $R_{\mathrm{QoS}}$. And an angular resolution $\Delta \theta_{\text{req}} = 15^\circ$ via (C1) in $\mathcal{P}1$. To counteract sensing uncertainties, the robust guard band confidence multiplier is set to $\kappa = 3$, guaranteeing a 99.73\% confidence level under the $3\sigma$ rule \cite{blackmore2011chance}.

\item \textit{Hardware Kinematic Constraints:} The PA's movement is bounded by a maximum velocity $v_{\mathrm{p}}^{\max} = 12.0$ m/s and an acceleration $a_{\mathrm{p}} = 980 \text{ m/s}^2$, reflecting the limits of industrial high-speed linear synchronous motors \cite{sato2015high}.
\end{itemize}

Next, we demonstrate three experiments and their results.

\subsection{Anti-blockage Movement of the PA}
\label{subsec: trajectory}
This experiment validates the anti-blockage movement of the PA under a simulated dynamic trajectory of the obstacle. 
The entire experiment lasts 30 seconds, comprising a total of 2702 frames. Fig.~\ref{Fig-experiment1-results} illustrates the spatial evolution of the PA's position, the user equipment (anchored at $X = 3.5$~m), and the moving mobile obstacle's trajectory. We can observe that the PA proactively performs the anti-blockage movement strategy as the moving obstacle approaches. Initially (before epoch 600), the PA maintains a stable LoS connection in \textbf{Mode 1 (Safety)}, as the unconstrained optimum position satisfies $x_{\mathrm{p}}^{\mathrm{opt}} \in \mathcal{X}_{\mathrm{safe}}$ near $X = 2.6$~m. As the obstacle approaches and $x_{\mathrm{p}}^{\mathrm{opt}}$ falls within the robust blockage zone $\tilde{\mathcal{Z}}_{\mathrm{b}}$, the system initiates \textbf{Mode 2 (Escaping)}. As depicted by the yellow trajectory in this mode, the PA first evades the obstacle by shifting to the valid extreme edges of $\mathcal{X}_{\mathrm{safe}}$ on this local side to determine the constrained optimal point $x_{\mathrm{p}}^*$. Subsequently, when the PA is pushed to the spatial extreme (around epoch 1000, $X \approx 10$~m) and $\tilde{\mathcal{Z}}_{\mathrm{b}}$ completely occupies the feasible interval on the current side, local boundary evaluation becomes invalid. Driven by the cycle, the PA must cross the robust blockage zone $\tilde{\mathcal{Z}}_{\mathrm{b}}$ to reach the newly determined $x_{\mathrm{p}}^* \in \mathcal{X}_{\mathrm{safe}}$ on the opposite side. After crossing, the PA resumes \textbf{Mode 1 (Safety)} and stabilizes at $x_{\mathrm{p}}^* = x_{\mathrm{p}}^{\mathrm{opt}}$. Based on the simulated mobile obstacle's trajectory in this experiment, the subsequent experiments will evaluate long-term system performance over 30 seconds, i.e., the full 2702 frames.

\subsection{Communication Performance under Comparative Baselines}
To verify the communication performance of our solution in maintaining effective throughput, three baseline comparative approaches (CA) are established below. These CAs have structural differences in terms of the effective radius, sensing ratio, and PA positioning strategy as summarized in Table~\ref{table_structures}. 
\begin{itemize}[leftmargin=*, nosep]
    \item \textit{Static Physical Blockage (CA1):} This scheme retains the joint optimization of the sensing ratio $\eta$ and the PA position $x_{\mathrm{p}}^*$, but disables the uncertainty and latency compensation mechanisms. It relies solely on the static physical boundary for anti-blockage movement (i.e., $\tilde{R}_{\mathrm{m}} = R_{\mathrm{m}}$). This configuration aims to isolate and validate the critical contribution of the formulated robust blockage zone $\tilde{\mathcal{Z}}_{\mathrm{b}}$ in combatting the unknown kinematics of indoor obstacles.
    
    \item \textit{Fixed Sensing (CA2):} This scheme employs the full dynamic effective radius $\tilde{R}_{\mathrm{m}}$ and optimizes the PA position $x_{\mathrm{p}}^*$, but omits the dynamic allocation of sensing resources by permanently fixing the sensing ratio at its predefined maximum (i.e., $\eta = \eta_{\max}$). This baseline highlights the fundamental necessity of the pre-optimized sensing parameters in balancing the sensing accuracy and the transmission window.
    
    \item \textit{Fixed PA (CA3):} This scheme serves as a conventional non-sensing baseline by fixing the PA at the center of the waveguide (e.g., $x_{\mathrm{p}} = 5.0$~m) and completely disables the ``sensing-movement-communication'' cycle (i.e., $\eta = 0$). It establishes the performance lower bound of traditional systems that lack anti-blockage movement capabilities, demonstrating their vulnerability to unknown indoor obstacles.
\end{itemize}

\begin{figure}[t]
	\centering
        \includegraphics[width=8.8cm]{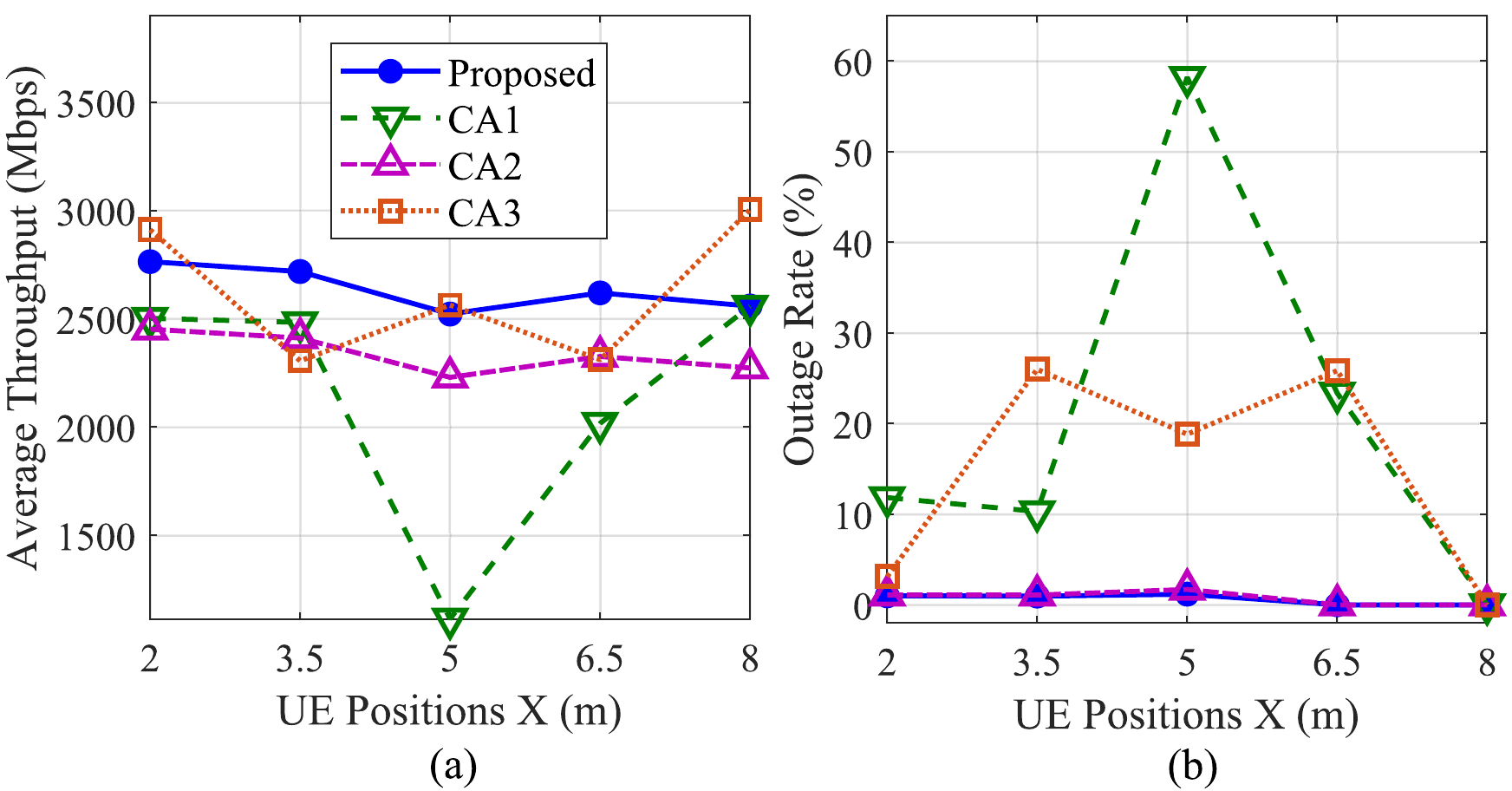}
    \caption{Long-term communication performance of our solution by averaging throughput and outage rate over 2702 running frames (in \cref{subsec: trajectory}) versus different user positions. }
    \label{Fig-experiment3-results}
\end{figure}

Fig.~\ref{Fig-experiment3-results} evaluates the communication performance of our solution against three benchmarks under varying initial user positions ($X_{\mathrm{user}} \in \{2.0, \allowbreak 3.5, \allowbreak 5.0, \allowbreak 6.5, \allowbreak 8.0\}$~m). The results demonstrate that our solution guarantees an outage rate of $0.9\%$ while sustaining a throughput range of 2850-3000~Mbps across all evaluated spatial scenarios. The Fixed PA scheme (CA3) achieves a higher throughput in non-obstructed edge scenarios (e.g., $X_{\mathrm{user}} = 2.0$ or $8.0$~m). This is because CA3 disables the ``sensing-movement-communication'' cycle ($\eta=0$), dedicating all temporal resources to data transmission. However, this temporal advantage introduces positional vulnerability. When the user is located at $X_{\mathrm{user}} = 3.5$~m, the mobile obstacle's trajectory eclipses the communication link, leading to an outage spike of $29.5\%$ and a drop in average throughput. This demonstrates that traditional fixed-antenna strategies are highly vulnerable to the unknown positions of blockages. Furthermore, despite possessing mechanical mobility, the Strict Physical Blockage Scheme (CA1) yields the highest outage rate (peaking at $64.1\%$) and a throughput degradation to $1100$~Mbps. This occurs because CA1 relies solely on raw physical boundaries without the robust margin $\tilde{R}_{\mathrm{m}}$, failing to absorb the tracking uncertainties of the obstacle's unknown kinematics. Alternatively, the Fixed Sensing Ratio Scheme (CA2) successfully avoids occlusions and maintains a low outage rate. However, because CA2 rigidly fixes the sensing ratio at its upper bound ($\eta_{\mathrm{max}}$) without using the pre-optimized sensing parameters, it excessively consumes temporal resources even when the blockage threat is low, resulting in a consistently lower throughput than our solution. 


\subsection{Impacts of System-level Parameters}
\begin{figure}[t]
	\centering
        \includegraphics[width=8.8cm]{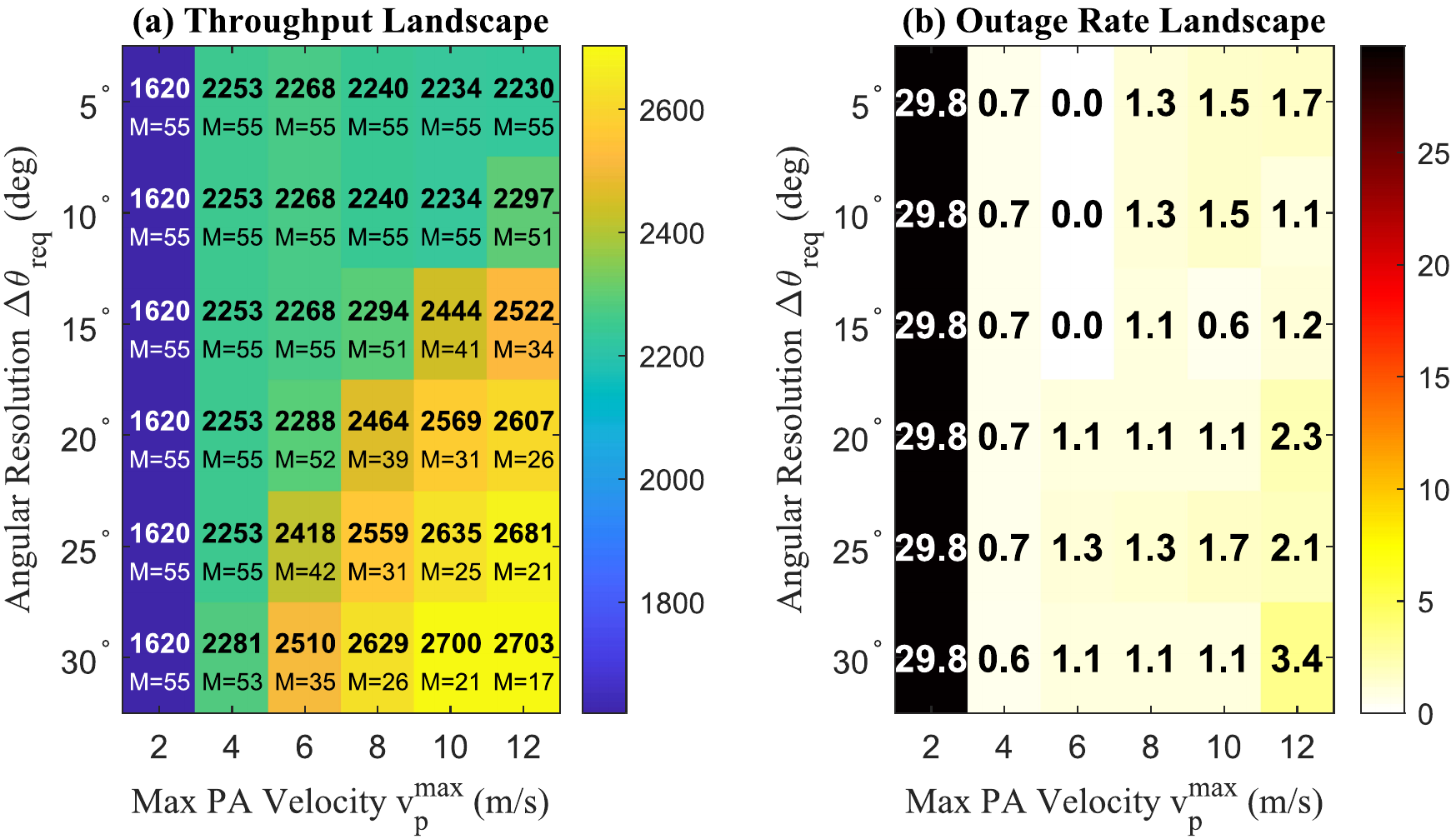}
    \caption{Impact of maximum mechanical speed $v_{\mathrm{p}}^{\max}$ and required angular sensing resolution $\Delta\theta_{\mathrm{req}}$ on the solution. }
    \label{Fig-experiment2-results}
\end{figure}

This experiment is to evaluate the performance impact of two system-level parameters: angular resolution $\Delta\theta_{\mathrm{req}}$ and maximum mechanical velocity $v_{\mathrm{p}}^{\max}$, under a fixed user position of $x_{\mathrm{u}} = 5$~m. As shown in Fig.~\ref{Fig-experiment2-results}, mechanical agility establishes the lower bound for movement safety, while sensing resolution restricts the communication capacity upper bound. This coupling is governed by the radar chirp count $M$. When $v_{\mathrm{p}}^{\max} \le 6$~m/s and $\Delta\theta_{\mathrm{req}} \le 10^\circ$, excessive sensing time ($M > 50$) bottlenecks capacity, causing up to a $17.1\%$ outage at $2$~m/s since the PA cannot outrun the occlusion. In the intermediate regime ($v_{\mathrm{p}}^{\max} \in [6, 10]$~m/s, $\Delta\theta_{\mathrm{req}} \in [15^\circ, 25^\circ]$), compressing the sensing ratio $\eta$ degrades signal quality and expands the robust blockage zone $\tilde{\mathcal{Z}}_{\mathrm{b}}$. This causes minor outage fluctuations ($\approx 1.0\%$), which are resolvable by elevating the velocity to $12$~m/s. However, extreme settings (e.g., $\Delta\theta_{\mathrm{req}}=30^\circ$ at $12$~m/s) drop the chirp count below the accuracy threshold ($M < 20$). This sensing failure renders spatial evasion infeasible and triggers a $5.6\%$ outage spike. For practical deployments, the system should operate beyond the kinematic bottleneck ($v_{\mathrm{p}}^{\max} \ge 6$~m/s) with a relaxed angular resolution ($\Delta\theta_{\mathrm{req}} \in [15^\circ, 25^\circ]$). Operating within this optimal regime ensures that the constrained optimization minimizes the sensing overhead $\eta$, translating the PA's mobility into sustained communication capacity for immersive applications.

\section{Conclusion}
This paper developed a sensing-assisted anti-blockage PASS solution to address the mmWave blockage vulnerability in indoor immersive communications. To overcome a critical limitation of unknown obstacle positions, our solution integrated FMCW radar sensing to PASS and executes a periodic ``sensing-movement-communication'' cycle. This solution enabled dynamic detection of mobile obstacles and proactively guides the PA to a safe zone (``escaping'') within the waveguide. Furthermore, we derived the pre-optimized sensing parameters to determine the efficient time allocation ratio and the scanning velocity of the PA during the sensing. Numerical results confirmed that our solution successfully minimizes temporal overhead, effectively suppressing the outage rate to near-zero levels while sustaining the pseudo-continuous, ultra-high capacity required by future 6G indoor immersive communication networks.

{\appendices


}

\bibliographystyle{IEEEtran}
\bibliography{references}

\end{document}